\documentclass{article}

\usepackage{times}
\usepackage{graphicx} 

\usepackage{caption}
\usepackage{subcaption}

\usepackage{natbib}

\usepackage{algorithm}
\usepackage{algorithmic}

\usepackage{hyperref}

\usepackage[accepted]{icml2016}

\usepackage{amssymb}
\usepackage{amsmath}
\DeclareMathOperator*{\argmax}{argmax}
\newcommand\cbr[1]{\left\{#1\right\}}
\newcommand\rbr[1]{\left(#1\right)}

\icmltitlerunning{Diversifying Music Recommendations}

\begin{document} 

\twocolumn[
\icmltitle{Diversifying Music Recommendations}

\icmlauthor{Houssam Nassif}{houssamn@amazon.com}
\icmlauthor{Kemal Oral Cansizlar}{kemal@amazon.com}
\icmlauthor{Mitchell Goodman}{migood@amazon.com}
\icmladdress{Amazon, Seattle, USA}
\icmlauthor{S.~V.~N.~Vishwanathan}{vishy@amazon.com}
\icmladdress{Amazon, Palo Alto, USA, and University of California, Santa Cruz, USA}

\icmlkeywords{Music, recommendation, submodular, Jaccard, diversity}

\vskip 0.3in
]

\begin{abstract} 
We compare submodular and Jaccard methods to diversify Amazon Music recommendations. Submodularity significantly improves recommendation quality and user engagement. Unlike the Jaccard method, our submodular approach incorporates item relevance score within its optimization function, and produces a relevant and uniformly diverse set.
\end{abstract} 

\section{Motivation}
With the rise of digital music streaming and distribution, and with online music stores and streaming stations dominating the industry, automatic music recommendation is becoming an increasingly relevant problem. Various recommender systems have been proposed, including models based on collaborative filtering~\cite{Xing2014banditCF}, content~\cite{Aaron2013deepMusic, Soleymani2015musicStructure}, context and emotions~\cite{Song2012musicRecSurvey}. Most of those recommender systems focus on improving recommendation accuracy and user preference modeling, in order to produce individually more enjoyable items. 

Unlike other digital and physical products, music content tends to have explicit clusters. An album contains multiple songs, all of which share the same album cover graphic, title and description. Furthermore, songs within the same album tend to belong to the same genre, and are usually played back to back. Due to their similar features, recommender systems tend to score same-album songs similarly.

It is common for music recommendations to be rendered in list form, which makes it easy for users to peruse on desktop, mobile or voice command devices. 
Naively ranking recommended songs by their personalized score results in lower user satisfaction because similar songs get recommended in a row.
Duplication leads to stale user experience, and to lost opportunities for music content providers wanting to showcase their content selection breadth. 
This impact is amplified on devices with limited interaction capabilities. For example, smart phones have a limited screen real estate, and it is usually more onerous to navigate between screens or even scroll down the page (see Figure~\ref{f:compare}). 

In fact, other factors besides accuracy contribute towards recommendation quality. Such factors include diversity, novelty, and serendipity, which complement and often contradict accuracy~\cite{Zhang2012Auralist}. Since we also deal with user-constructed libraries, we focus on exploring methods to diversify music recommendations. 

This work presents experiments to diversify Amazon Music mobile app recommendations. Amazon offers Prime members a free Prime Music benefit, with access to millions of songs and thousands of expert-programmed playlists. Customers can also upload their own music to their library, and mix it with Prime Music content to create personal playlists. Amazon Music developed a recommender system that assigns a personalized score to each music content.

\section{Diversity Methods}
Similar to cases
in visual discovery~\cite{Teo2016airstream}, image search~\cite{van2009visual}, blog posts~\cite{elarini2009turning}, and news articles~\cite{ahmed2012fair}, we
apply diversification to alleviate recommendation redundancy. We consider two different diversity methods, one based on Jaccard distance, and the other on submodularity.

\subsection{Jaccard Swap Diversity}
The Jaccard distance measures dissimilarity between two finite sample sets $A$ and $B$:
\begin{equation}\label{e:jaccard_distance}
d_J(A,B) = \frac{|A \cup B| - |A\cap B|}{|A \cup B|}.
\end{equation}
For each candidate music recommendation, we use an explanation-based diversification method to generate a set of weighted corresponding explanatory items~\cite{Yu2009algoSwap}. The explanatory items are latent features of the candidate, generated from content and behavioral features. We compute the Jaccard distance between two music recommendations by applying Equation~\ref{e:jaccard_distance} to their underlying explanatory items~\cite{Clarkson2006Survey}. We generate a list of $k=40$ recommendations using the Algorithm Swap method~\cite{Yu2009algoSwap}, which iteratively maximizes top $k$ pair-wise Jaccard distance, conditioned on score relevance.

\subsection{Submodular Diversity}
Alternatively, we formulate the selection and ranking of a diverse musical subset as a submodular optimization problem~\cite{fujishige2005submodular}. Submodular functions are characterized by a diminishing returns property. For set $S$, subset $A \subseteq S$, elements $x,y \in S$, and submodular function $f:\{0,1\}^S \rightarrow \mathbb{R}$, we have:
\begin{equation}\label{e:diminishingReturns}
f(A\cup\{x\}) - f(A) \geq f(A\cup\{x,y\}) - f(A\cup\{y\}).
\end{equation}
We divide all musical content into $C$ categories, according to the same content attributes. Each scored candidate gets mapped to multiple categories based on its content and behavioral features. Each category $c$ has its own submodular function $f_c$. To ensure that a candidate contributes no more than its personalized score, we use:
\begin{equation}
f_c(A) = \log \rbr{1+\sum_i score(a_i)}.
\end{equation}

We diversify by maximizing the sum $\rho$ of all category functions $f_c$, which is itself submodular: 
\begin{equation}\label{e:rho}
\argmax_A \rbr{\rho(A) = \sum_c^C f_c(A)}.
\end{equation} 
A near-optimal solution is achieved by an iterative greedy procedure~\cite{nemhauser1978analysis}:
\begin{equation}\label{e:greedy}
  A_0 := \emptyset ~\text{and}~ 
  A_{i+1} := A_i \cup \cbr{\argmax\limits_{a \in A \backslash
      A_i} \rho(A_i \cup \cbr{a})}.
\end{equation}

\section{Results and Discussion}
\begin{figure}[ht]
  \centering
  \begin{subfigure}[t]{0.22\textwidth}
    \centering
    \includegraphics[width=1\textwidth]{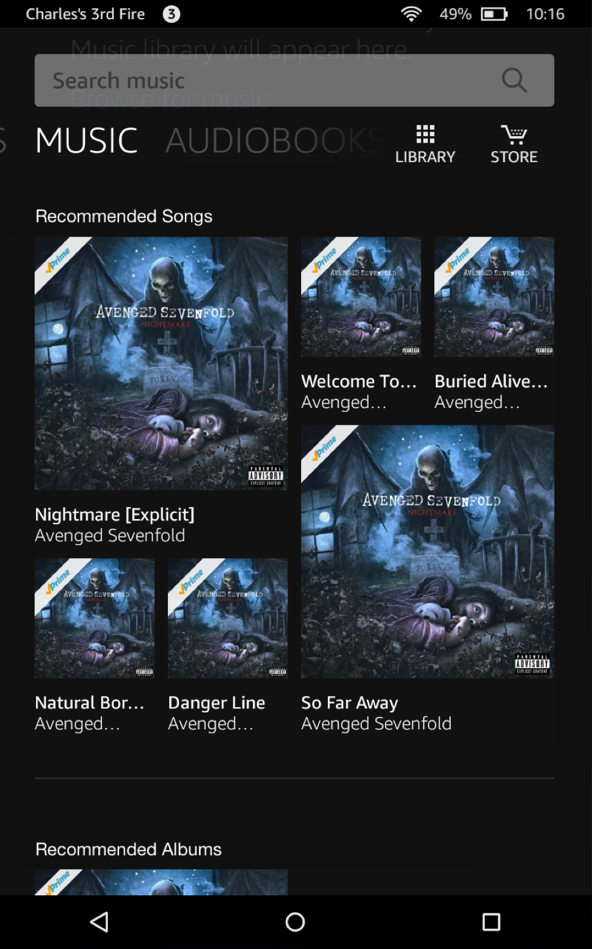}
    \caption{Baseline recommendation}
  \end{subfigure}
  \quad
  \begin{subfigure}[t]{0.22\textwidth}
    \centering
    \includegraphics[width=1\textwidth]{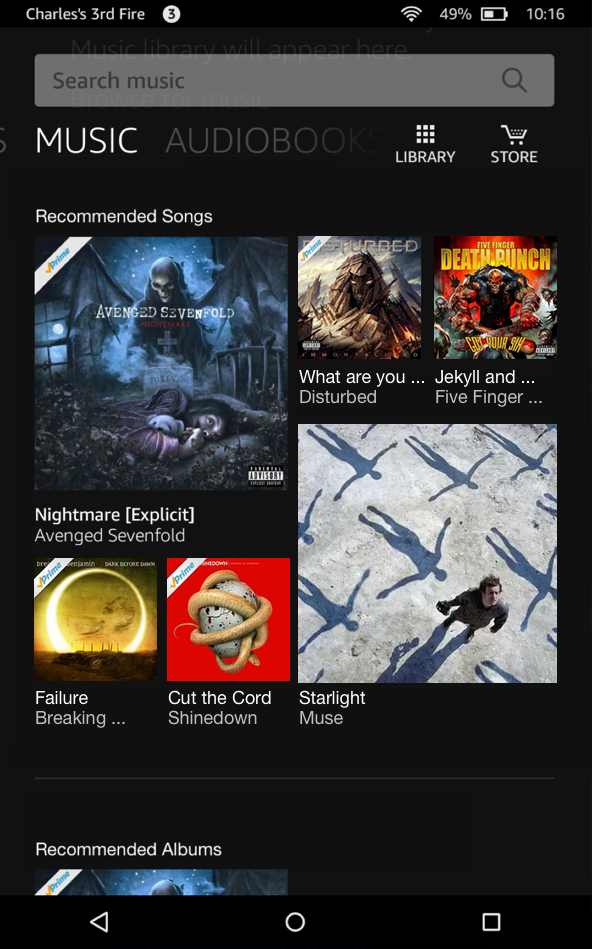}
    \caption{Submodular diversity}
  \end{subfigure}
  \caption{Effect of diversification on Amazon Prime Music mobile app personalized album recommendations.}
  \label{f:compare}
\end{figure}

We test the effectiveness of our diversity methods using an online experiment to improve customer engagement with the Amazon Prime Music app track, album and playlist recommendations (Figure~\ref{f:compare}). We run a 3-way A/B test on top of the Amazon Music recommender system: baseline, Jaccard Swap diversity, and submodular diversity. Recommender uses item-to-item collaborative filtering and provides item score and explanatory set~\cite{linden2003Sims}. We use the artist and album features as attributes/categories for the Jaccard/submodular methods. Baseline ranks by recommender score and lacks diversity. The experiment lasted 4 weeks, with equal allocation of at least 700,000 customers per treatment. We evaluate the treatment impact on engagement by tracking the number of minutes streamed. We compare the method's lift~\cite{nassif2013Uplift} via Welch's t-test. 

\vskip -0.1in
\begin{table}[ht]
\caption{Increase in number of minutes streamed.}
\label{t:results}
\begin{center}
\begin{small}
\begin{tabular}{lcc}
\hline
\abovespace\belowspace
Treatment & Jaccard Swap & Submodular\\
\hline
\abovespace
Baseline & $0.40\%$ $(p=0.18)$ & \boldmath{$0.64\%$ $(p=0.03)$} \\
\belowspace
Jaccard Swap &            & $0.24\%$ $(p=0.41)$ \\
\hline
\end{tabular}
\end{small}
\end{center}
\vskip -0.1in
\end{table}

Table~\ref{t:results} shows experimental results. Based on minutes streamed, both diversity measures fare better than baseline. This result reinforces the notion that diversity affects recommendation quality~\cite{Zhang2012Auralist}. Only submodularity's $0.64\%$ lift improvement is significant (Figure~\ref{f:compare}).

Our submodular solution outperforms the Jaccard approach. One reason may be that maximizing the submodular function $\rho$ (Equation~\ref{e:rho}) by iteratively picking the category with the highest gain (Equation~\ref{e:greedy}) produces a uniformly diverse stream, as the diminishing returns property holds for any contiguous part of the recommended list. Jaccard Swap doesn't guarantee such a smooth list. 

Another possible reason is due to Algorithm Swap, which does not necessarily retain the most relevant content at the head of the list. The swap can sacrifice a highly relevant song in order to increase overall diversity~\cite{Yu2009algoSwap}. The submodular approach ensures that the most relevant song appears first, followed by a mix of the most relevant songs within each category.

\bibliography{musicDiversity}
\bibliographystyle{icml2016}

\end{document}